\documentstyle[12pt]{article}
\begin{document}
\rightline{Preprint BROWN-HET-1029, CLNS 95/1383, DAMTP 95-73}
\begin{flushright}
{hep-ph/9601327, January 1996}
\end{flushright}
\vspace*{3mm}
\begin{center}
{\bf The Impossibility of Baryogenesis at a Second Order 
Electroweak Phase Transition 
}
\end{center}
\bigskip
\centerline{\large\rm Tomislav Prokopec\footnote{\rm e-mail: 
tomislav@hepth.cornell.edu}}
\begin{center}
{\it Newman Laboratory for Nuclear Studies, Cornell University, 
Ithaca NY 14853, USA}
\end{center}
\centerline{\large\rm Robert Brandenberger\footnote{\rm e-mail:
rhb@het.brown.edu}}
\begin{center}
{\it Physics Department, Brown University, Providence RI 02912, USA}
\end{center}
\begin{center}
and
\end{center}
\centerline{\large\rm Anne-Christine Davis
\footnote{\rm e-mail: a.c.davis@damtp.cam.ac.uk}}
\begin{center}
{\it DAMTP, University of Cambridge, Cambridge CB3 9EW, UK}
\end{center}

\begin{center}

{\large\bf Abstract}

\end{center}

We investigate whether baryogenesis is possible 
at a second order electroweak phase transition.
We find that under rather general conditions, 
the departure from thermal equilibrium is suppressed
by the expansion rate of the Universe, and hence baryon 
production is also suppressed by the expansion rate. 
We conclude that if no additional sources of departure from thermal
equilibrium such as topological defects are present, then electroweak
baryogenesis is ruled out if the phase 
transition is second order or crossover. However, a non-vanishing net 
baryon to
entropy ratio {\it is} generated, and we provide both upper and lower 
bounds
for the result. 
Our technique is also applicable to other baryogenesis mechanisms
taking place during and immediately after a second order phase 
transition.
We estimate the lowest value of the transition energy scale for which 
the
resulting baryon to entropy ratio might be large enough to explain 
observations.



\section{Introduction}

Recently the constraints on the Higgs sector, at least in the Minimal 
Standard Model have improved so that the prevalent view today
is that the electroweak phase transition is either very weakly first 
order, second order, or perhaps even crossover. The first evidence 
is
a direct lower limit for the mass of the minimal Higgs particle
$M_H>65$GeV (95\% C.L.) set in experiments
at CERN \cite{CERN} (for a review, see for example \cite{CERNreview}) 
and
based on the lack of Higgs production in hadronic decays of the Z 
boson. 
An indirect limit arises from the precision electroweak data;
when the CDF and D0 measurements of the top quark mass are taken 
account
of ($m_t\simeq 181\pm 12 $GeV), one arrives at the $1\sigma$ allowed
range for the Minimal Standard Model (MSM) Higgs mass: 
$26$GeV$<M_H<230$GeV
($68\%$ C. L.) \cite{EllisFogliLisi}. In addition the requirement 
for (meta)stability
of the electroweak vacuum \cite{AltarelliIsidori}, 
\cite{CasasEspinosaQuiros} implies a lower bound of $M_H>116$GeV,
and the requirement that the Standard Model couplings remain 
perturbative 
up to a scale $\Lambda_p\sim M_{Pl}\sim 10^{19}$GeV results in 
a perturbative upper bound $M_H\leq 190$GeV \cite{Sher}. Taking into 
account all of the above we see that there are strong
indications that the MSM Higgs mass is between 100 and 200 
GeV.

In theories with an extended Higgs sector
(with two or more Higgs doublets) the experimental limits on the 
mass of the lightest Higgs scalar are somewhat weaker. 
For example, in the Minimal Supersymmetric Standard Model (MSSM)
the direct limits from the LEP experiments
\cite{CERNmssm}
on the lightest neutral Higgs boson mass are roughly $M_h>40$GeV.
The direct constraints on the mass of the second neutral Higgs boson
$M_A$ become rather weak  once the radiative corrections from other 
supersymmetric particles of unknown mass are taken into account. 
The meta-stability bounds do not apply to the MSSM vacuum, while the 
intrinsic upper bound on $M_h$ is about $140$GeV (for $m_t\leq 
190$GeV)
\cite{CasasEspinosaQuirosRiotto}.
So, in the case of the MSSM, the bounds on the lightest Higgs mass
indicate a somewhat lighter Higgs particle: $40$GeV $< M_h < 140 
$GeV. 

Next we discuss how the nature of the electroweak phase transition 
depend
on the details of the Higgs sector, and in particular on the mass of 
the
(lightest) Higgs particle. 
We now understand the nature of the phase 
transition in the Minimal Standard Model for a moderately light
Higgs $M_H\leq m_W\simeq 81 $GeV. The pictures of the phase 
transition 
emerging from two-loop calculations
\cite{ArnoldEspinosa}, \cite{FodorBuchmuller} and lattice
calculations \cite{KajantieLRS} agree and indicate that the 
transition 
is strongly first order for small Higgs mass and becomes weaker as 
the mass becomes comparable to $m_W$. It is believed that 
for a sufficiently large Higgs mass the phase transition
eventually becomes second order
\cite{ArnoldYaffe}, \cite{BuchmullerPhilips}, \cite{Wetterich}
or even a crossover, since
at large Higgs masses the theory simply resembles
more and more scalar theory. At what Higgs mass this exactly happens 
is not clear at the moment. Also how the dynamics of
the phase transition is affected by the presence of
more than one Higgs doublet, as is the case in the supersymmetric 
versions of the Standard Model, is not completely understood. 
It is possible, for example, that the phase transition
proceeds in two stages \cite{LandCarlson}, but we will not consider 
this possibility here. We are mainly interested in how the strength 
of the 
phase transition changes; even though there is no complete analysis 
of a general supersymmetric two Higgs doublet model, in the case of 
the MSSM it is known \cite{BrignoleEspinosaQuirosZwirner}
that the stop sector tends to strengthen and a light CP odd neutral
Higgs scalar tends to weaken the phase transition. 
We conclude that, given the above considerations, it is  likely 
that the electroweak phase transition is second order or crossover.

In this letter we will address the question: 
{\it Is baryogenesis at a second order electroweak phase 
transition ruled out?\/} (For reviews of electroweak baryogenesis
see \cite{CKNreview}, \cite{Turokreview}.) We will assume that no 
other
sources of departure from thermal equilibrium are present
except for those caused by the phase transition and the 
expansion rate of the Universe. This means, for example, that 
we will not consider the case when one has a network of 
cosmological defects \cite{BDPT} 
floating around that could drive the system out of equilibrium. 

The letter is organized as follows. In Section 2 we derive a simple
dissipative equation of motion which out of equilibrium may lead to 
a net baryon number production. In the subsequent section 
we analyze this equation, derive an upper bound on the number of 
baryons
produced and find that it is suppressed as the expansion rate of the 
Universe.
In Section 4 we discuss a lower bound on the strength of baryogenesis 
at a
second order phase transition and use the result to estimate the 
minimal 
energy scale for a second order baryogenesis mechanism without extra 
out 
of equilibrium effects such as topological defects. 
In the final section we summarize our results and their consequences 
for
electroweak baryogenesis.

\section{Equation of motion}

We start from a standard near equilibrium formula of statistical 
mechanics for the rate of approach to equilibrium of some charge $Q$ 
\begin{equation}
\dot Q=-\Gamma_Q {\Delta F\over T} 
\label{eq:dot Q}
\end{equation} 
where $\mu_Q=\Delta F$, the chemical potential for $Q$,
measures by how much the free energy changes when $Q$ changes by one
unit, and 
$\Gamma_Q$ is the rate of decay of $Q$ (per unit time).
This equation is a macroscopic form of the
charge (non)conservation and can be thought of as an integral of  
the Boltzmann equation (see for example 
\cite{Balescu}).

When applied to the baryon number ($Q=B$), we get
(see \cite{DLSFP} for a similar analysis):
\begin{equation}
\dot {n_B}\equiv \frac{\dot B}{V}
=-\frac{\Gamma_{sph}}{V} {\Delta F_{(\Delta B=1)}\over T} \Delta B 
(sph)\,,
\quad \Delta F_{(\Delta B=1)} =\mu_B
\label{eq:dot baryon}
\end{equation} 
$\dot B$ is the total rate of change of baryon number,
$\Gamma_{sph}/V$ the sphaleron rate per unit volume, 
$T$ is the temperature of the plasma,
$\mu_B=\Delta F _{(\Delta B=1)}$ the chemical potential for baryon
number (which we defined as the change in free energy when 
baryon number changes by one unit), and 
$\Delta B (sph)$ is the change in baryon number per sphaleron 
transition.
%
%
In the symmetric phase the sphaleron rate is
$\Gamma_{sph}/V=\kappa (\alpha_w T)^4$ with  $\kappa\simeq 1.1$ 
\cite{ambjornkrasnitz}, \cite{philipsen},
and in the broken phase 
\begin{equation}
\frac{\Gamma_{sph}}{V} \propto \exp{-{{E_{sph}} \over T}},
\label{eq:sphrate}
\end{equation}
where $E_{sph}$ is the sphaleron energy. 

We will now illustrate how using Eq. (\ref{eq:dot baryon}) and a one 
loop
formula for the free energy of fermions in a thermal plasma one can 
obtain 
an equation for near equilibrium  baryon number production.
The (one-loop) thermal contribution to the free energy reads:
\begin{equation}
F=T\sum_i\int {d^3 p\over (2\pi)^3}
\ln \left ( 1+\exp (-\beta (E_i-\mu_i))\right ),
\label{eq:free energy}
\end{equation} 
where the {\it sum} $\sum_i$ is over all fermionic degrees of freedom 
and 
the energy in the presence of a non-vanishing $\dot\theta$ is 
given (in the WKB limit) by \cite{JoyceProkopecTurokII}
\begin{equation}
E_i= \left [ (|\vec p |\mp g_\theta \dot\theta )^2 +m_i^2 \right ]^{1/2}\qquad 
{\rm for} \quad \Sigma^3=\pm 1
\label{eq:energy}
\end{equation} 
where $g_\theta =-v_2^2/2 (v_1^2+v_2^2)$ is the `$\theta$-charge,'
(which is related to the axial charge), 
$v_1$ and $v_2$ are the vacuum expectation values ({\it vev\/}s)
of the two Higgs fields, respectively, 
and $\Sigma^3$ is the spin of the particle. 
In the zero mass limit the particles with 
$\Sigma^3=+1$ reduce to the left handed fermions or left handed 
anti-fermions (anti-particles of the right handed fermions),
whilst those with $\Sigma^3=-1$ reduce to the right handed fermions 
or right 
handed anti-fermions. Even though this relation is derived using 
equilibrium techniques and $\dot\theta$ is a time dependent quantity, 
we expect it to be a good approximation to description of near
equilibrium processes since the typical time scale at which 
$\dot\theta$
changes is given by the expansion rate of the Universe and 
the processes which equilibrate the system are typically much faster. 

For any particle species denoted by a subscript $i$, it follows from 
Eqs.~(\ref{eq:dot Q}) and (\ref{eq:dot baryon}) that the number 
density
$n_i$ obeys the following equation
({\it cf.\/} \cite{JoyceProkopecTurokI}):
\begin{equation}
\dot n_i=-{\Gamma_{sph}\over VT} \nu_i \sum_j\nu_j\mu_j
\label{eq:dot ni}
\end{equation} 
where $\nu_j$ are the stechiometric coefficients of the reaction and 
$\mu_j$ are the corresponding chemical potentials. In the case of 
sphalerons, 
$\Gamma_{sph}/2V$ is the rate per unit volume for each of the 
following two sphaleron transitions:
$t_L t_L b_L \tau_L ...\leftrightarrow 0$,
$t_L b_L b_L \nu_\tau ...\leftrightarrow 0$, where {\it dots\/} 
($...$) 
denote the particles of the lighter two families. The coefficients 
$\nu_i$ 
of the first process are for example $\nu_{t_L}=2$,  $\nu_{b_L}=1$, 
{\it etc,\/} so that (\ref{eq:dot ni}) for the left-handed top quarks
and left-handed bottom quarks yields:
\begin{eqnarray}
\dot n_{t_L} & = &-{\Gamma_{sph}\over 2VT}
[5\mu_{t_L} + 4 \mu_{b_L} +2\mu_{\tau_L}+\mu_{\nu_\tau}\\
\nonumber
& & +5\mu_{c_L}+4 \mu_{s_L} +2\mu_{\mu_L}+\mu_{\nu_\mu}
+5\mu_{u_L}+4 \mu_{d_L} +2\mu_{e_L}+\mu_{\nu_e} ]\\
\dot n_{b_L} & = & -{\Gamma_{sph}\over 2VT} 
 [4\mu_{t_L} + 5 \mu_{b_L} +\mu_{\tau_L}+2\mu_{\nu_\tau}\\
\nonumber
& & +4\mu_{c_L}+5 \mu_{s_L} +\mu_{\mu_L}+2\mu_{\nu_\mu}
+4\mu_{u_L}+5 \mu_{d_L} +\mu_{e_L}+2\mu_{\nu_e} ]
\label{eq:dot tLbL}
\end{eqnarray} 
Having in mind that the total baryon number density $B$ is
\begin{equation}
B = \frac{1}{3}\sum_{\rm quark\;\; species}  n_i,
\label{eq:baryon number}
\end{equation}
and $\dot n_{i_R}=0$ for right handed particles, one gets:
\begin{eqnarray}
\dot B & = & - N_F{\Gamma_{sph}\over 2VT} 
[3\mu_{t_L} + 3 \mu_{b_L} +\mu_{\tau_L}+\mu_{\nu_\tau}\\
\nonumber
& & +3\mu_{c_L}+3 \mu_{s_L} +\mu_{\mu_L}+\mu_{\nu_\mu}
+3\mu_{u_L}+3 \mu_{d_L} +\mu_{e_L}+\mu_{\nu_e} ]
\label{eq:dot B}
\end{eqnarray} 
where $N_F=3$ is the number of families. Note that we have ignored the
anti-particles; it is rather trivial to include them in 
Eq. (\ref{eq:dot B}): one should just subtract the chemical 
potentials for the
anti-particles. 

The final step is to relate
the chemical potentials in this relation to the number densities.
The particle number density can be easily obtained from 
Eq. (\ref{eq:free energy}) as follows:
\begin{equation}
n_i=-\frac{\partial}{\partial E_i}F=
\int {d^3 p\over (2\pi)^3}\frac{1}{1+\exp [\beta (E_i-\mu_i) ]}
\label{eq:number density}
\end{equation} 
Next we expand the number density to linear order in $\mu_i$:
%
\begin{eqnarray}
n_i & = & n_i^0+\beta\int {d^3 p\over (2\pi)^3}
\frac{\exp (\beta E_i^0)}{(1+\exp (\beta E_i^0))^2}
\left [ - E (p,\dot\theta, m_i)+ E(p,0,m_i)+\mu_i \right ]\;\;\;
\nonumber\\
n_i^0 & = &
\int {d^3 p\over (2\pi)^3}\frac{1}{1+\exp (\beta E_i^0)}
\,,\qquad E_i^0=({p^2+m_i^2})^{1/2}.
\label{eq:number density expanded}
\end{eqnarray} 
After some algebra (see Appendix A) we get (to leading order in 
$\dot\theta$
and in the high temperature limit $T> m_i$)
\begin{equation}
\mu_i  = \frac{12}{T^2} [ (n_i-n_i^0)\pm c(m_i^2) g_\theta\dot\theta 
]\,,\qquad
c(m_i^2) =  \frac{1}{4\pi^2} m_i^2(1 - {m_i \over {3 T}}).
\label{eq:chemical potential}
\end{equation} 
We have displayed the cubic mass term just to get a feeling when the 
high temperature expansion breaks down; for $m_i\simeq T$ we expect 
it still 
to be reasonably accurate. Eq.~(\ref{eq:dot B}) can now be re-written 
as
\begin{equation}
\dot B=- 6 N_F{\Gamma_{sph}\over VT^3} 
\sum_i \left [ n_{iL} \pm c (m_i^2) g_\theta\dot\theta 
\right ] 
\label{eq:dot baryon II}
\end{equation} 
where $N_F=3 $ is the number of families, 
$\sum_i n_{iL} = n_L$ is the total left handed fermion number 
density, which can be  recast in terms of baryon and lepton
numbers as $3B_L+L_L$.
(In the above the equilibrium contribution from particles 
$\sum_i n_i^0$ has been cancelled by the contribution from 
anti-particles 
$-\sum_i \bar n_i^0$.) We write the final form of the baryon number 
equilibration equation as follows:
\begin{equation}
\dot B=- 6 N_F{\Gamma_{sph}\over VT^3} \left [ 
3 B_L+ L_L + 6 c (m_t^2) g_\theta\dot\theta
\right ] 
\label{eq:dot baryon III}
\end{equation} 
where the contribution from the light fermions to $c(m_i^2)$ has been 
ignored 
in comparison to the top quark. The coefficient $6=2N_c$ in front of 
the $\dot\theta$ term is the 
number of 
degrees of freedom for the left-handed top quark. 

In the next section we discuss the solution to this equation assuming 
that the 
phase transition is second order or crossover. 

\section{An upper bound on the baryon to entropy ratio}

We will now analyze Eq.~(\ref{eq:dot baryon III}) and, assuming that 
before the transition the Universe is in thermal equilibrium so 
that baryon number is {\it zero\/} initially, we will investigate 
what 
is the maximum baryon number produced at a second order 
phase transition in a two Higgs doublet model.

First we simplify Eq.~(\ref{eq:dot baryon III}). 
Since the time scale for the weak sphaleron transitions 
is at least $\tau_{sph}\sim 1/(\alpha_w^4 T) $ , $\tau_{sph}$ is 
much larger than the time scale of strong sphaleron transitions 
$\tau_{ss}\sim 1/(\alpha_s^4 T) $, so that at any moment when
discussing only the dynamics of the weak sphaleron processes,
the strong sphalerons will be to a very good approximation 
in chemical equilibrium, which means $B_L=B_R$. 
We also have to make an assumption about the time scale for
equilibration of the left and right lepton numbers 
$\tau_{LR}\sim 1/(\alpha_w y_l^2 T)$  (here $y_l\sim y_\tau\sim 
10^{-2}$
is the Yukawa coupling constant). 
For simplicity we will assume $\tau_{LR}<< \tau_{sph}$, which 
implies that to a good approximation $L_L=L_R$. This will certainly 
be overwhelmingly satisfied when the weak sphaleron rate is 
exponentially suppressed, the case of most interest to us.
When the symmetry is restored and when the sphaleron rate is
un-suppressed, 
$\tau_{LR}> \tau_{sph}$ may be a more reasonable approximation.
We will not study this case here since, as we  will see below,
most of the baryons are produced at a time when the sphaleron
rate is exponentially suppressed. In addition
we know that $B-L$ is conserved by the the Standard Model. 
For definiteness we will set $B-L$ to zero, which
is motivated by the symmetric initial conditions. 
To summarize we have:
\begin{eqnarray}
B_L & =B_R \qquad  & \tau_{sph}>> \tau_{ss}\\
\nonumber
B & =L\qquad & ({\rm symmetric\;\; initial\;\; condition})\\
\nonumber
L_L & =L_R\qquad & \tau_{sph}>> \tau_{LR}
\nonumber
\end{eqnarray}
>From these constraints we easily infer $B_L=L_L=B/2$ which implies
$3B_L + L_L = 2B$. Eq.~(\ref{eq:dot baryon III}) can be now
written in a simple form
\begin{eqnarray}
\dot B=- \Gamma (B+{\cal C} \dot\theta)\;,\quad  
{\cal C} = g_\theta \frac{3 m_t^2}{2\pi^2} (1 - {{m_t} \over {3 
T}})\,,
\quad \Gamma= 12 N_F \frac{\Gamma_{sph}}{V T^3} ,\quad  
\end{eqnarray}
where $m_t= y_t\phi$ is the top mass. 
Here, $\phi$ is the Higgs expectation value which changes during the 
electroweak phase transition from $\phi = 0$ to 
$\phi = \phi(T=0)$.

The solution to this equation with the zero initial baryon number can 
be 
easily obtained by the Greens function method 
\begin{equation}
B(t)=-\int_{t_{in}}^t \, dt'\, {\cal C}\dot\theta (t') \Gamma (t')\,
\exp [-\int _{t'}^t \Gamma (t '') dt ''].
\label{eq:baryon number expanded}
\end{equation}

In order to derive an upper bound on $B(t)$, we write
\begin{equation}
B(t) \leq {\cal B}_0 \int_{t_{in}}^t d{t^\prime}\Gamma({t^\prime})
\exp{[-\int_{t^\prime}^t d{t^{\prime\prime}} 
\Gamma(t^{\prime\prime})]} 
={\cal B}_0 
\label{eq: new1}
\end{equation}
with
\begin{equation}
{\cal B}_0 = {\rm max}_{{t^\prime} \in [t_{in}, t]} |C({t^\prime}) 
{\dot \theta}({t^\prime})|
\label{eq: new2}
\end{equation}
and we have assumed that $\int _{t_{in}}^{t_f}\Gamma dt >>1$. 
This formula is our upper bound for baryon number production.
Next we will show that, in the context of a simple model 
for the effective potential near the phase transition, ${\cal B}_0$ 
is suppressed by the expansion rate of the Universe. We believe
that our argument is not limited to the simple form for the potential
we use:
\begin{equation}
V(\phi, T) = D (T^2-T_0^2)\phi^2 +\frac{\lambda}{4} \phi^4
\label{eq: effective potential}
\end{equation}
Here we quote the one-loop Standard Model values for the parameters:
$T_0^2\simeq 1.39 M_H^2+ (60 GeV)^2$ is the spinodal temperature,
$D\simeq 0.18$, and $\lambda$ depends on the Higgs mass 
($M_H^2/2v_0^2$, $v_0=246$GeV) and, in addition, 
it is weakly (logarithmically) dependent on temperature.
$\lambda\sim 0.05 - 0.16$, depending on the Higgs mass,
the lower bound corresponding to $M_H = 80$GeV,
the upper to $M_H = 140$GeV \cite{DineLeighHuetLindeLinde}.
The value of $\lambda$ is weakly dependent on temperature;
for example when temperature decreases by about $10\%$ (which will
turn out to be the value at which (\ref{eq: new2}) is maximized), the
value of $\lambda$ changes by not more than $10\%$, which we can 
ignore.
Since we do not trust the one loop values at the above quoted Higgs
masses, we will not discuss the values of these parameters in 
the MSSM. The reader should keep in mind that 
we do not know what the true form of the effective potential is
near the phase transition, since we are interested in the case of a 
large Higgs mass for which the perturbative expansion breaks down.
A more appropriate treatment may be the $\epsilon$-expansion as 
advocated in \cite{ArnoldYaffe} or some other technique based on 
renormalization group \cite{Wetterich}.
However, it is not clear whether the $\epsilon$-expansion 
is accurate either in the case of the Standard Model or its 
supersymmetric version, so we do not know very well what the form 
of the effective potential is. The non-perturbative calculations
\cite{KajantieLRS}, \cite{Wetterich} are still not at 
the level to be able to reconstruct the form of the potential for 
large Higgs masses.
In order to illustrate our mechanism we made a simple assumption
on the form of the effective potential, which is
motivated by the one-loop finite temperature effective potential
(with the cubic term, which gives rise to 
a first order phase transition, set to zero).
This form of the effective potential should 
give a reasonable qualitative description the phase transition
since it has the desirable generic features of a second order phase 
transition: it is maximally flat at the critical temperature where the
quadratic term vanishes; the order parameter $\phi$ changes 
continuously as temperature drops. We believe that the use of the
true form of the effective potential in the analysis that follows
would not alter our main observation that ${\cal B}_0$
is suppressed by the expansion rate of the Universe at all times. 

The time dependence of $C(t)$ and ${\dot \theta}(t)$ in (\ref{eq: 
new2}) are both determined by $\phi(t)$, the value of the Higgs 
scalar expectation value. 
In Appendix B we show that to a very good approximation
$\phi(t)$ is given by the location 
of the minimum of the finite temperature effective potential
(\ref{eq: effective potential}):
%
\begin{equation}
\phi_0^2(T) = {{2 D} \over \lambda} [T_0^2 - T^2] \qquad {\rm for} 
\quad T < T_0.
\label{eq: new3}
\end{equation}
The relative departure of the true  {\it vev\/} from this value, as 
is shown
in Appendix B, is suppressed by the expansion rate of the
 Universe. This remains true when the friction of the $\Phi$ field 
is taken into account.

In order to obtain the time dependence of $\theta$, we make the
simplified assumption
\begin{equation}
\theta\equiv \epsilon\frac{\phi(T)}{\phi (T=0)}
=\epsilon [1-T^2/T_0^2]^{1/2},
\label{eq: theta}
\end{equation}
where $\epsilon$ is the net change in the CP violating phase 
(the relative phase between the Higgs doublets in the case of the two 
Higgs doublet model).
Cline {\it et al\/} \cite{ClineKainulainenVisher}
have done an extensive study
of the dynamics of $\theta$ and $\phi$ fields neglecting friction
and find that for rather large range of parameters the above linear
approximation is reasonable. As the analysis of Appendix B suggests, 
this conclusion should not be altered even when the friction of both
fields is taken into account. 

In order to relate time derivatives to the derivatives with respect 
to 
temperature we will use the Friedmann equation for the radiation era:
\begin{equation}
H=\frac{1}{2t}= h \sqrt {g_*}\frac{T^2}{m_{Pl}}\,,\qquad
h^2=\frac{8\pi}{3}\frac{\pi^2}{30}\,,\quad G=m_{Pl}^{-2}\,,\quad
g_*=106.25
\label{eq: Hubble parameter}
\end{equation}
so that $tT^2=\, const$ and $d/dt=-HT d/dT$, or $Hdt=-dT/T$.

>From Eqs.~(\ref{eq: theta}) and~(\ref{eq: Hubble parameter}) it 
then follows that
\begin{equation}
{\dot \theta} = H \theta {1 \over { ({T_0/ T})^2 - 1}}
= \epsilon H {T \over T_0} \left [ {T_0^2 / T^2} - 1 
\right]^{-1/2},
\label{eq: new4}
\end{equation}
and thus, maximizing ${\cal C}\dot\theta$, gives
\begin{equation}
{\cal B}_0 = {24 \over {25\sqrt 5 \pi^2}}| g_\theta \epsilon|
 y_t^2 {{2 D} \over \lambda} H (T_0) T_0^2.
\label{eq: new5}
\end{equation}

Using the equation
\begin{equation}
s = {{2 \pi^2} \over {45}} g_* T_0^3
\label{eq: new6}
\end{equation}
for the entropy density,  the baryon to entropy ratio is bounded from 
above by 
\begin{equation}
{B \over s} \leq {{108} \over {5\sqrt 5 \pi^4}} {y_t^2 \over g_*} 
| g_\theta\epsilon | 
{{2 D} \over \lambda} {H (T_0) \over T_0}.
\label{eq: new7}
\end{equation}
For $T_0 = 110$GeV ($M_H=80$GeV), the value of $H / T_0$ 
is about $1.6 \cdot 10^{-16}$, and for 
$T_0 = 175$GeV ($M_H=140$GeV), $H / T_0\simeq 2.5 \cdot 10^{-16}$;
$g_\theta\simeq -1/4$ (we have assumed here that $v_1=v_2$), 
$y_t\simeq 1/\sqrt 2$.
$ 2 D / \lambda$ is bounded from above by about $7$ 
when  $M_H=80$GeV. (For a larger value $M_H=140$GeV, 
$ 2 D / \lambda\simeq 2.5$.) 
Our final bound for the baryon to entropy ratio is then:
\begin{equation}
{B \over s} \leq 1.5\times 10^{-19} |\epsilon | ,
\label{eq: new8}
\end{equation}
which completes the proof that without additional sources of 
out-of-thermal
equilibrium such as topological defects, electroweak baryogenesis is
not possible if the electroweak phase transition is not first order.

\section{A lower bound on the baryon to entropy ratio}

We are now interested in obtaining an approximate lower bound on the 
baryon number today (which corresponds to $B=B(\infty)$).  
It proves convenient to split the integral in 
Eq.~(\ref{eq:baryon number expanded})
into two parts: the first for which the sphaleron rate $\Gamma$ is 
large
so that the time integral in the exponent is much larger than one, and
the second part for which this integral is smaller than one. The time 
$t_H$ which separates the two regimes is defined by 
\begin{equation}
\int _{t_H}^\infty \Gamma (t '') dt ''=1.
\label{eq: integral t H}
\end{equation}

On the first interval $[t_{in}, t_H]$ we can perform a partial 
integration to obtain
\begin{equation}
\nonumber
B(\infty)=-{\cal C}\dot\theta (t_H) {\rm e}^{-1}
+{\cal C}\dot\theta (t_{in}){\rm e}^{ - \int _{t_{in}}^{t_H}\Gamma 
(t) dt}{\rm e}^{-1}
\nonumber
\end{equation}
\begin{equation}
+\int_{t_{in}}^{t_H} \, dt'\frac{d({\cal C}\dot\theta )}{dt'} 
{\rm e}^{-\int _{t'}^{t_H} \Gamma (t '') dt ''}{\rm e}^{-1}
-\int_{t_{H}}^\infty \, dt'\, {\cal C}\dot\theta (t') \Gamma (t')\,
{\rm e}^{ -\int _{t'}^\infty \Gamma (t '') dt ''}
\label{eq: baryon number in pieces}
\end{equation}
We can immediately see that the second term vanishes 
if $t_{in}$ is chosen such that the Universe is initially in the
symmetric phase.

The idea behind the derivation of the lower bound is the following: 
the
first and fourth terms have a negative sign, the third term is 
positive.
We will show that the third term is smaller than the first in absolute
value, and that hence the absolute value of the baryon number can be
bounded from below by the absolute value of the final term.

To find an upper bound on the third term, we replace the exponential 
factor
by $1$. The remaining integral exactly cancels the first term. Thus, 
the
absolute value of $B(\infty)$ can be bounded from below by a lower 
bound on
the absolute value of the last term.

In order to bound the last term, it is necessary to know the exact 
temperature dependence of $\Gamma$ (recall that for obtaining the 
upper bound on $B(t)$
this was not required). In the case of sphaleron-induced baryogenesis 
at the electroweak scale, we have:
\begin{equation}
\Gamma(T) = \gamma T y^7 e^{-By}, \,\,\,\, y = \sqrt{{{4 \pi} \over 
{\alpha_w}}}
{\phi \over T},
\label{eq:new13}
\end{equation}
where $B \sim 1.5 - 2.7$ is a slowly varying function of the coupling 
constant ratio, and $\gamma \sim 10^{-4} \kappa_1$, where 
$\kappa_1$ is the one loop fluctuation determinant.   
Before we discuss the last term in 
(\ref{eq: baryon number in pieces}),
we will evaluate the integral in Eq. (\ref{eq: integral t H}).
We can solve this integral approximately by
using the equation (\ref{eq:new13})  for $\Gamma$ 
and expressing the time differential in terms of the temperature
differential with the help of the Friedmann equation 
(\ref{eq: Hubble parameter}). The result is 
\begin{eqnarray}
\frac {\Gamma(x_H)}{T(x_H)} & \simeq & \frac{2D}{\lambda}
\frac{B}{x_H}
\sqrt{\frac{4\pi}{\alpha_w}}\, \frac{H(T_0)}{T_0}
\\ \nonumber
x_H\equiv \frac{\phi_H}{T} & =  &
\frac{1}{B(g^2/\lambda )}\sqrt{\frac{\alpha_w}{4\pi}}
\ln\left [
\gamma \frac{\lambda}{2D} \left (\frac{4\pi}{\alpha_w}\right )^3
\frac{1}{B(g^2/\lambda)}
\frac{T_0}{H(T_0)}x_H^8
\right ]
\end{eqnarray}
which is solved for $x_H=\phi_H/T\simeq 1.2$.
This is nothing but the condition for the sphaleron erasure bound of 
Ref.~\cite{Shaposhnikov} (see below). 
 
With the help of Eqs.~(\ref{eq:chemical potential}), 
(\ref{eq: new4}) and (\ref{eq: integral t H}), the last term of 
Eq.~(\ref{eq: baryon number in pieces}) can now be estimated
as follows: 
\begin{eqnarray}
& &|\int_{t_{H}}^\infty \, dt'\, {\cal C}(t')\dot\theta (t') \Gamma 
(t')\,
{\rm e\/}^{ -\int _{t'}^\infty \Gamma (t '') dt ''}|
> {\rm e}^{-1}| \int_{t_{H}}^\infty \, dt'\, 
{\cal C}(t')\dot\theta (t') \Gamma (t')|
\\ \nonumber
& &
\simeq {\rm e}^{-1}{3 \over {2 \pi^2}}\left ( \frac{2D}{\lambda} 
\right ) ^{1/2}
|\epsilon\, g_\theta| y_t^2 
\frac{x_H}{[x_H^2 (\lambda/2D) +1 ]^3}
{H(T_0) T_0^2}
\simeq \frac{{\cal B}_0}{{\rm e}}\,.
\label{eq: lower bound on baryon number}
\end{eqnarray}
Comparing with Eq.~(\ref{eq: new5}) it follows that the lower bound 
on the
baryon to entropy ratio produced in the second order phase transition 
is 
only suppressed by  a factor of ${\rm e}^{-1}$ compared to the upper 
bound.

\section{Conclusions}

In this paper we have shown that without additional sources which 
drive the system out of thermal equilibrium (such as a network of 
topological defects),
baryogenesis during a second order electroweak phase transition is 
much too weak to be able to produce the observed baryon to entropy 
ratio. 

We have shown that rather generically baryon number production in 
this case is suppressed 
by the expansion rate  of the Universe. The case studied was that 
of a two Higgs doublet model with CP violation in the Higgs sector with
the assumption that both Higgs fields couple the same way to the plasma
and hence the phase transition occurs simultaneously for both Higgs
fields. Even though in principle it would be interesting to study a 
more generic situation of a two stage phase transition, we believe
that the main conclusions of our letter still hold. 

In general, the source of CP violation may not be 
in the Yukawa terms, but in the neutralino and 
chargino mass matrices \cite{HuetNelson}.
In this case the effective shift in free energy should be
proportional to a time derivative of the imaginary part of the masses,
{\it i.e.\/} a gauge or Higgs coupling  times a time derivative of the 
relative Higgs phase, so that the analysis is very similar
to the one presented in this letter and again baryon production is
suppressed by H. 

The second main result is that a non-vanishing baryon asymmetry
 {\it is} generated during
such a phase transition. We have derived upper and lower bounds for
the strength of this mechanism. They are of the same order
 of magnitude and suppressed by a factor
$H / T_0$, where $H$ is the Hubble expansion rate and $T_0$ is
the temperature of the phase transition. Note that the baryon 
production mechanism presented in this paper does not suffer from 
the sphaleron erasure problem 
\cite{Shaposhnikov} since the bulk of baryons is produced
when the sphaleron rate is switching off and becomes smaller than the
expansion rate of the Universe.

We have illustrated our mechanism for the effective potential which 
describes a second order phase transition, such that 
the order parameter changes continuously at the phase 
transition but its (time) derivative is discontinuous, as opposed
 to crossover when the order 
parameter is a smooth function of time. However, from the point of
view of baryon production in our mechanism, it is irrelevant whether
the transition is second order or crossover since the bulk of baryon
production occurs sufficiently below the phase transition when the
difference between these two types of phase transition becomes 
immaterial.

The formalism illustrated in this paper may be applicable to other 
baryogenesis mechanisms taking place in a second order phase 
transition. 
Provided that a formula similar to Eq. (\ref{eq:dot baryon III}) 
holds, 
then an analysis similar to what was done in this paper would be
applicable. The most optimistic value of the resulting $B / s$ is
$\epsilon H / T_0 \sqrt{g_*}\sim \epsilon T_0 / m_{pl}$,
where as before
$\epsilon$ is a constant parameterizing the strength of CP violation 
in
the particle physics sector determining the phase transition. Hence, 
provided that the scale of symmetry breaking is higher than about
$10^{9}$GeV \footnote{A more realistic
lower bound on the scale of symmetry breaking
is about $5\times 10^{10}$GeV;
this is easily obtained from Eq.~(\ref{eq: new8}) and the requirement 
$\epsilon \leq 1$.}
from the lower bound formula, it is possible to imagine a 
sufficiently powerful
baryogenesis scenario involving only second order phase transitions.

\section{Appendix A: Derivation of the expression for the chemical 
potential}

In this Appendix we outline the derivation of 
Eq.~(\ref{eq:chemical potential}) starting from
Eq.~(\ref{eq:number density}) in the limit
$\dot\theta \rightarrow 0$ and  $m_i < T$.

Linearizing in $\dot\theta$, Eq. (\ref{eq:number density}) gives
\begin{equation}
n_i = n_i^0 + {{\mu T^2} \over {2 \pi^2}} I_\mu \mp {{g_\theta 
\dot\theta T^2}
\over {2 \pi^2}} I
\label{eq:a1}
\end{equation}
where
\begin{equation}
I_\mu = \int_0^\infty x^2 dx  
{{\exp{\sqrt{x^2 + m_T^2}}} \over {(1 + \exp{\sqrt{x^2 + m_T^2}})^2}}
\label{eq:a2}
\end{equation}
and
\begin{equation}
I = \int_0^\infty {{x^3 dx} \over {\sqrt{x^2 + m_T^2}}} 
{{\exp{\sqrt{x^2 + m_T^2}}} \over {(1 + \exp{\sqrt{x^2 + m_T^2}})^2}}.
\label{eq:a3}
\end{equation}
The notation $m_T = m_i / T$ and $x = p / T$ has been used.

The integral $I_\mu$ can be performed explicitly in the limit $m_T 
\rightarrow
0$, yielding $\zeta_2=\pi^2 / 6$. The second term is more tricky since
the leading term
(which is independent of $m_T$) just represents a shift in $\mu$. The 
method 
of evaluating $I$ consists of breaking up the integration region into 
two
intervals, the first from $0$ to $y$, where $m_T \ll y \ll 1$, and 
the second
from $y$ to $\infty$. In the first interval, we can set the 
exponential 
factors to 1 and evaluate the remaining integral to leading order in 
$m_T$, in
the second interval we introduce a new integration variable $z = 
\sqrt{x^2 + m_T^2}$, obtaining
\begin{equation}
I_2 = \int_{\sqrt{y^2 + m_T^2}}^\infty dz (z^2 - m_T^2) 
{\exp{z} \over {(\exp{z} + 1)^2}}.
\label{eq:a4}
\end{equation}
The second integral (coefficient $m_T^2$) can be performed explicitly 
and is 
of the order $m_T^2$, the first can be performed by first integrating 
from 
$0$ to $\infty$ and then subtracting the result obtained by 
integrating from 
$0$ to $\sqrt{y^2 + m_T^2}$. The first can be done explicitly, and 
gives a
constant $\zeta_2$ which can be viewed as a shift in $\mu$, the second
integral can be 
approximated to leading order in $\sqrt{y^2 + m_T^2}$, and to this 
order 
exactly cancels the leading $y$ dependence of integration of $I$ from 
$0$
to $y$. A good computational check is that the final result must
be independent on the cut-off $y$, since the cut-off is introduced 
solely
for computational convenience.
Thus, to leading order in $m_T$, the result for $I$ is 
\begin{equation}
I = \zeta_2 - {{m_T^2} \over 2} ( 1 - {{m_T} \over 3}).
\label{eq:a5}
\end{equation}
%
The first term in $I$ should have no physical effect
since in the zero mass limit $\dot \theta$ is pure gauge and must 
have no
physical effect whatsoever. It can be viewed as a shift in the 
chemical
potential $\mu$: 
\begin{equation}
\bar\mu = \mu \mp g_\theta \dot\theta,
\label{eq:a6}
\end{equation}
where $\bar\mu$ denotes the physical chemical potential with a
correct physical interpretation:  $\bar\mu$ is  
proportional to the particle density perturbation.
With the substitution of Eq. (\ref{eq:a6}), and inserting the above 
results for $I$ and $I_\mu$, Eq. (\ref{eq:a1}) yields the result
for $\bar \mu$ given in Eq. (\ref{eq:chemical potential}).
(Note that in the main text we have denoted the physical (shifted)
chemical potential by $\mu$.)

\section{Appendix B: The irrelevance of dissipative effects}

In this Appendix we will show that in the case of a second order 
phase 
transition dissipation does not effect the evolution of the $\phi$ 
field 
and it is irrelevant for baryogenesis considerations. This is 
in contrast to the case of a first order phase 
transition where it is known \cite{MooreProkopec} that dissipation 
is crucial for the dynamics of the transition and baryogenesis. We 
will 
study the case of the $\phi$ field, but an analogous analysis applies 
to 
the relative Higgs phase $\theta$. 

The equation of motion for the Higgs condensate is 
\begin{equation}
\ddot\phi+3H\dot\phi+\gamma_\phi\dot\phi + 
\frac{\partial V(\phi, T)}{\partial \phi} = 0.
\label{eq: phi evolution}
\end{equation}
The form of the dissipative term $\gamma_\phi\dot\phi$ can be 
obtained 
in the WKB approximation 
using the method developed in \cite{MooreProkopec}. Assuming that 
the main source of dissipation comes from the tree-level process in 
which
the Higgs particle scatters off the top and decays, we obtain for the 
friction  
\begin{equation}
\gamma_\phi\simeq \frac{3}{2\pi^4}\lambda\phi^2\ln 
\frac{T^2}{\lambda\phi^2} \frac{1}{\Gamma_\phi},
\end{equation}
where $\Gamma_\phi\simeq 0.3 \alpha_w y_t^2 T$ is the Higgs decay 
rate.
The lessons to learn from this simple calculation are the following.
Firstly, the friction term is local in time. This is to
be expected since the time scale on which the Universe evolves $\sim 
1/H$
is huge in comparison to the time scale on which perturbations are 
damped $\tau_\phi\sim 1/\Gamma_\phi$.
This is in contrast to what happens at a first order phase transition 
where
the system is perturbed out of equilibrium in a non-local manner in 
the
sense that perturbations are sourced and then propagate across the 
phase boundary. 
Secondly, in the case of one fluid, friction is related to viscosity,
while in our case of many fluids interacting, the main source of
friction are couplings between different species.
Thirdly, the friction for the $\phi$ field is {\it immense\/} in 
comparison to the expansion rate of the Universe. 

Since the time 
scale on which the Universe evolves is given by $\sim 1/H$, we can 
neglect 
the first two terms in Eq.~(\ref{eq: phi evolution}). 
We now write Eq.~(\ref{eq: phi evolution}) in the following form: 
\begin{equation}
\phi=\phi_0+\varphi\,,\qquad 
\gamma_\phi \dot\phi
+
8D(T_0^2-T^2)\varphi +3 [ 2D\lambda (T_0^2-T^2) ]^{1/2}\varphi ^2
+\lambda\varphi^3
=0
\label{eq: phi evolution II}
\end{equation}
The solution is
\begin{equation}
\varphi=-\phi_0\frac{\gamma_\phi H}{T^2}\frac{1}{8D}
\frac{1}{(T_0^2/T^2-1)^2}
\end{equation}
which means that the instantaneous value of $\phi$ is very well 
approximated
by the equilibrium value $\phi_0$ in the sense that 
the relative correction $\varphi/\phi$
is suppressed by the expansion rate of the Universe. 
One can check that indeed all of the following
$\dot \varphi$, $\varphi^2$ and $\varphi^3$ 
are suppressed by at least $H$ in comparison to $\dot\phi_0$, which 
justifies the above approximate solution. 

We have thus shown that friction does not alter in any significant 
manner
the evolution of the Higgs condensate and the use of $\phi_0$ in the
main text was justified. An analogous consideration applies to the
$\theta$ field.

\section{Acknowledgments}

We acknowledge funding from the U.K. PPARC (TP and ACD), the U.S. NSF
(TP) and the U.S. DOE under Grant DE-FG02-91ER40688-Task A (RB). 
We are grateful to Michael Joyce and Neil Turok for useful 
discussions.

\end{document}